\documentstyle[twoside,fleqn,espcrc2,epsf]{article}

\makeatletter
\def\lesssim{\mathrel{\mathpalette\vereq<}}
\def\vereq#1#2{\lower3pt\vbox{\baselineskip1.5pt \lineskip1.5pt
\ialign{$\m@th#1\hfill##\hfil$\crcr#2\crcr\sim\crcr}}}

\def\alt{\lesssim}

\makeatother

\begin{document}
\thispagestyle{empty}

\title{
\vbox to0pt
{\vskip0pt minus100cm\centerline{\fbox{\parbox{\textwidth}
{\small\noindent Proc.\ AXION WORKSHOP,
University of Florida, Gainesville, Florida, USA, 13--15 March
1998, ed.\ by P.~Sikivie, to be published in 
Nucl.\ Phys.\ B Proc.\ Suppl.}}}
\vskip2.2cm}\vskip-22pt\noindent
Stellar-Evolution Limits on Axion Properties}

\author{Georg~Raffelt\\
        {\ }\\
        Max-Planck-Institut f\"ur Physik, F\"ohringer Ring 6, 
        80805 M\"unchen, Germany}

%\date{\today}

\begin{abstract}
If axions exist, they are copiously produced in hot and dense plasmas,
carrying away energy directly from the interior of stars. Various
astronomical observables constrain the operation of such anomalous
stellar energy-loss channels and thus provide restrictive limits on
the axion interactions with photons, nucleons, and electrons. In
typical axion models a limit $m_a\alt 10^{-2}~{\rm eV}$ is
implied. The main arguments leading to this result are explained,
including more recent work on the important supernova 1987A
constraint.
\end{abstract}

\maketitle

%%%%%%%%%%%%%%%%%%%%%%%%%%%%%%%%%%%%%%%%%%%%%%%%%%%%%%%%%%%%%%%%%%%%%%
%% Section I %%%%%%%%%%%%%%%%%%%%%%%%%%%%%%%%%%%%%%%%%%%%%%%%%%%%%%%%%
%%%%%%%%%%%%%%%%%%%%%%%%%%%%%%%%%%%%%%%%%%%%%%%%%%%%%%%%%%%%%%%%%%%%%%

\section{INTRODUCTION}

\vfil

In the late 1950s, the existence of a direct neutrino-electron
interaction was postulated in the context of the universal $V{-}A$
theory of weak interactions. It was immediately recognized that it
enables the production of neutrinos in stellar plasmas by thermal
processes of the sort $\gamma e^-\to e^- \bar\nu\nu$ or plasmon decay
$\gamma\to\bar\nu\nu$. As neutrinos escape unscathed from normal
stars, such processes effectively constitute a local energy sink for
the nuclear reactions which power stars. Neutrino losses soon became a
standard ingredient of stellar evolution theory~\cite{Clayton} even
though the details had to be modified when additional neutrino flavors
and neutral-current interactions appeared.

New low-mass particles or nonstandard neutrino couplings would
increase ``invisible'' stellar energy losses and thus modify the
observed properties of stars, notably the duration of certain
evolutionary phases. This ``energy-loss argument'' was first applied
in 1963 to constrain neutrino dipole moments which would lead to
nonstandard neutrino losses through an enhancement of the plasmon
decay rate $\gamma\to\bar\nu\nu$ \cite{Bernstein}.  A first
application to ``exotic'' particles appeared in 1975 to constrain the
coupling strength of putative light Higgs bosons~\cite{Sato}. Since
then a large variety of cases has been studied, where axions were
perhaps the greatest motivation for a systematic refinement of this
method~\cite{Raffelt90,RaffeltBook}.

The purpose of this overview is to explain the stellar energy-loss
argument in the context of the most important cases, to summarize the
axion limits obtained from this method, and to point at some open
issues that require further research.

%%%%%%%%%%%%%%%%%%%%%%%%%%%%%%%%%%%%%%%%%%%%%%%%%%%%%%%%%%%%%%%%%%%%%%
%% Section I %%%%%%%%%%%%%%%%%%%%%%%%%%%%%%%%%%%%%%%%%%%%%%%%%%%%%%%%%
%%%%%%%%%%%%%%%%%%%%%%%%%%%%%%%%%%%%%%%%%%%%%%%%%%%%%%%%%%%%%%%%%%%%%%

\begin{figure}[b]
\hbox to\hsize{\hss\epsfxsize=2.5cm\epsfbox{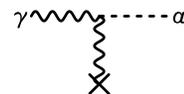}\hss}
\caption{\label{figprimakoff} Primakoff production of axions.}
\end{figure}

\section{SUN}

\begin{figure}[ht]
\hbox to\hsize{\hss\epsfxsize=6cm\epsfbox{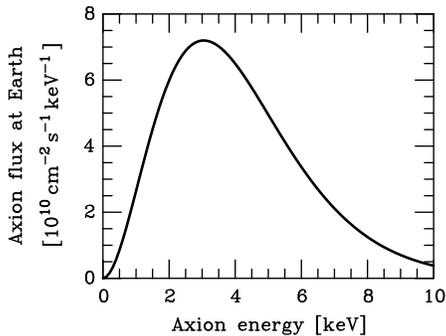}\hss}
\caption{\label{figsunax} Solar axion flux at Earth from the
  Primakoff conversion of thermal photons in the Sun for
  $g_{a\gamma}=10^{-10}~{\rm GeV}^{-1}$~\protect\cite{Bibber}.}
\end{figure}

While the Sun does not provide the most restrictive astrophysical
axion limits, it remains of interest as a source for
previous~\cite{Lazarus}, current~\cite{helioscope1} or
proposed~\cite{helioscope2} experiments to search for the solar axion
flux. Of particular interest is the axion-photon interaction
\begin{equation}
{\cal L}_{a\gamma}=g_{a\gamma}{\bf E}\cdot{\bf B}\,a,
\end{equation}
where $g_{a\gamma}$ is a coupling constant with the dimension
(energy)$^{-1}$, ${\bf E}$ and ${\bf B}$ are the electric and magnetic
fields, respectively, and $a$ is the axion field. This interaction
allows for the Primakoff conversion of photons into axions and vice
versa (Fig.~\ref{figprimakoff}) and thus serves both as a production
process in the Sun and as a detection process in ``helioscope''
experiments where one attempts to back-convert solar axions into
detectable x-rays either in a dipole magnet or in a
crystal~\cite{Lazarus,helioscope1,helioscope2}. The expected solar
axion spectrum is shown in Fig.~\ref{figsunax} for a coupling strength
$g_{a\gamma}=10^{-10}~{\rm GeV}^{-1}$.

The ``standard Sun'' is about halfway through its main-sequence
evolution. Therefore, the solar axion luminosity must not exceed
its photon luminosity or else its nuclear fuel would have been spent
before reaching the current age of $4.5\times10^9~{\rm yr}$. This
simple requirement yields~\cite{Dearborn87}
\begin{equation}\label{eq:solarlimit}
g_{a\gamma}\alt 24\times10^{-10}~{\rm GeV}^{-1}.
\end{equation}
Interestingly, even such a large axion luminosity could be
accommodated in a solar model by a suitable adjustment of the unknown
presolar helium abundance.

The recent advance in helioseismology allows one to derive sound-speed
profiles throughout the Sun---it is no longer enough for a solar model
to reproduce the observed luminosity and radius at an age of
$4.5\times10^9~{\rm yr}$. We are currently investigating if and how
much one can improve on Eq.~(\ref{eq:solarlimit}) by using
helioseismological observations as a solar-model
constraint~\cite{Schlattl}.

%%%%%%%%%%%%%%%%%%%%%%%%%%%%%%%%%%%%%%%%%%%%%%%%%%%%%%%%%%%%%%%%%%%%%%
%% Section II %%%%%%%%%%%%%%%%%%%%%%%%%%%%%%%%%%%%%%%%%%%%%%%%%%%%%%%%
%%%%%%%%%%%%%%%%%%%%%%%%%%%%%%%%%%%%%%%%%%%%%%%%%%%%%%%%%%%%%%%%%%%%%%

\begin{figure}[ht]
\hbox to\hsize{\hss\epsfxsize=\hsize\epsfbox{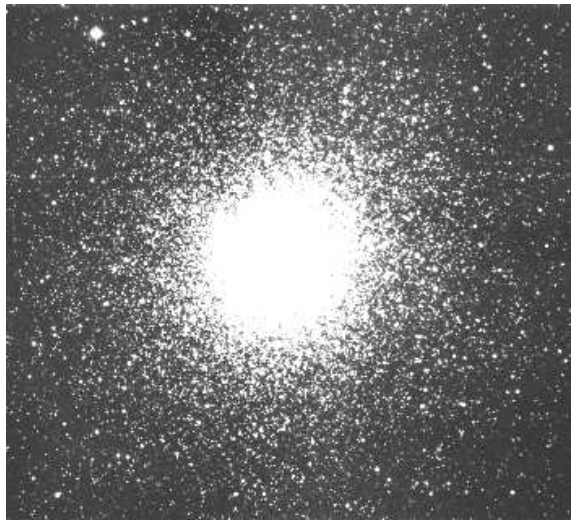}\hss}
\caption{\label{figm3} Globular cluster M3.}
\end{figure}

\section{GLOBULAR CLUSTERS}

The most restrictive constraint on the axion-photon coupling as well
as on a possible interaction with electrons arises from
globular-cluster stars. Our galaxy has about 150 globular clusters
such as M3 (Fig.~\ref{figm3}), each consisting of a gravitationally
bound system of typically $10^6$ stars. The galactic globular-cluster
system forms a roughly spherical halo---they are not located in the
disk.  The escape velocity from a typical cluster is only around
$10~{\rm km~s^{-1}}$ so that a single supernova is enough to sweep it
clean of all gas, preventing further star formation and thus
guaranteeing almost equal stellar ages. In addition, all stars of a
given cluster have nearly equal metallicities, where ``metals'' in
astrophysical parlance are the elements heavier than helium.
Therefore, the stars of a given cluster differ primarily in a single
parameter, the initial mass, providing an ideal ensemble to test the
theory of stellar evolution.

\begin{figure}[b]
\hbox to\hsize{\hss\epsfxsize=\hsize\epsfbox{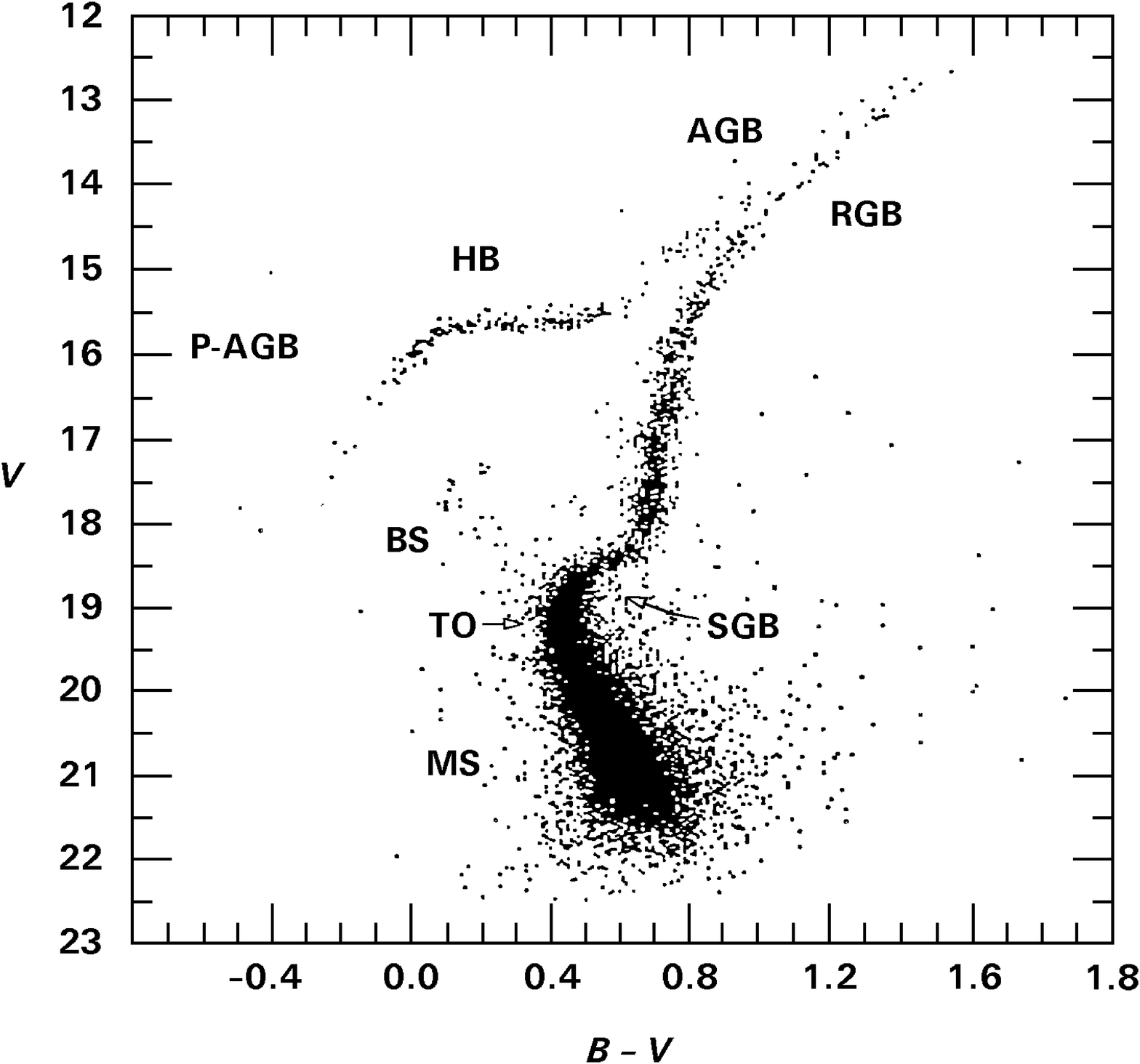}\hss}
\vskip-6pt
\caption{\label{figcolmag}Color magnitude diagram for the globu\-lar
cluster M3 \protect\cite{Buonanno86}, based on the photometric data of
10,637 stars.  The classification for the evolutionary phases is as
follows~\protect\cite{Renzini}.  MS (main sequence): core hydrogen
burning. BS (blue stragglers). TO (main-sequence turnoff): central
hydrogen is exhausted.  SGB (subgiant branch): hydrogen burning in a
thick shell. RGB (red-giant branch): hydrogen burning in a thin shell
with a growing core until helium ignites. HB (horizontal branch):
helium burning in the core and hydrogen burning in a shell. AGB
(asymptotic giant branch): helium and hydrogen shell burning. P-AGB
(post-asymptotic giant branch): final evolution from the AGB to the
white-dwarf stage.}
\end{figure}

\begin{figure}[b]
\hbox to\hsize{\hss\epsfxsize=\hsize\epsfbox{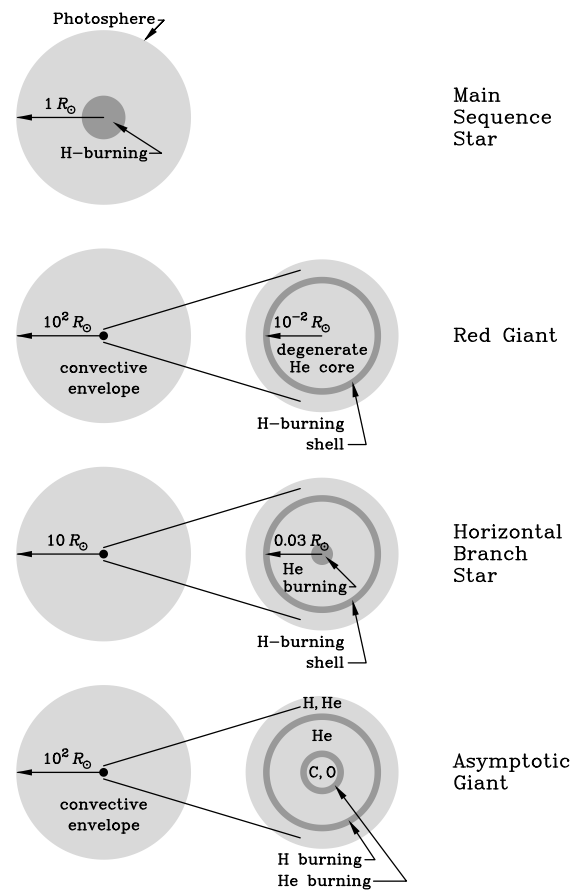}\hss}
\vskip-12pt
\caption{\label{figphases} Evolutionary phases of low-mass
  stars. The envelope and core dimensions depend on the location on
  the RGB, HB, or AGB, respectively; the radii are given for
  a rough orientation.}
\end{figure}

To this end one arranges the stars of a cluster in a color-magnitude
diagram where one plots the color, representative of surface
temperature, on the horizontal axis and the brightness on the vertical
axis.  One thus obtains a characteristic pattern
(Fig.~\ref{figcolmag}) where different branches correspond to
different evolutionary phases which are schematically illustrated in
Fig.~\ref{figphases}.

The main sequence (MS) corresponds to central hydrogen burning and
thus to normal stars like our own Sun. When the central hydrogen fuel
is exhausted, the stars develop a degenerate helium core, with
hydrogen burning continuing in a shell. The luminosity is governed by
the gravitational potential at the edge of the growing helium
core---these stars become ever brighter as they ascend the red-giant
branch (RGB). The higher a star is on the RGB the more massive (and
compact) its helium core has become. Curiously, the envelope of these
stars expands, leading to a large surface area and thus a low surface
temperature (red color)---hence the name ``red giant.''

The helium core grows until it reaches about $0.5\,M_\odot$ (solar
mass) when it becomes dense and hot enough to ignite helium.  The
ensuing core expansion reduces the gravitational potential at the edge
and thus lowers the energy production rate in the hydrogen shell
source. Helium ignition dims these stars, even though they now have
two energy sources!  The core masses are equal when helium ignites,
but the envelope mass can differ due to varying rates of mass loss on
the RGB, leading to different surface areas and thus surface
temperatures.  Therefore, after helium ignition the stars occupy the
horizontal branch (HB) in the color-magnitude diagram.

Finally, when helium is exhausted, a degenerate carbon-oxygen core
develops, leading to a second ascent on what is called the asymptotic
giant branch (AGB). These low-mass stars cannot ignite their CO core.
They become white dwarfs after shedding most of their envelope.

The advanced evolutionary phases are fast compared with the MS
evolution of about $10^{10}~{\rm yr}$ for stars with a total mass
somewhat below $1\,M_\odot$. For example, the ascent on the upper RGB
and the helium-burning phase each take around $10^8~{\rm yr}$.
Therefore, the distribution of stars along the RGB and beyond can be
taken as an ``isochrone'' for the evolution of a single star, i.e.\ a
time-series of snapshots for the evolution of a single star with a
fixed initial mass.  Put another way, the number distribution of stars
along the different branches are a direct measure for the duration of
the advanced evolutionary phases. The distribution along the MS is
different in that it measures the distribution of initial masses.

The core temperature before and after helium ignition is about
$10^8~{\rm K}$, but the average core density changes from
$2\times10^5~{\rm g~cm^{-3}}$ \hbox{(degenerate)} to
$0.6\times10^4~{\rm g~cm^{-3}}$ on the HB (nondegenerate).  This
implies that the Primakoff production of axions will be much more
effective in the cores of HB stars than in those of upper RGB stars.
The axionic energy loss on the HB would shorten the helium-burning
phase because the nuclear fuel would be consumed faster, causing a
reduction of the number of HB relative to RGB stars.

\begin{figure}[t]
\hbox to\hsize{\hss\epsfxsize=6cm\epsfbox{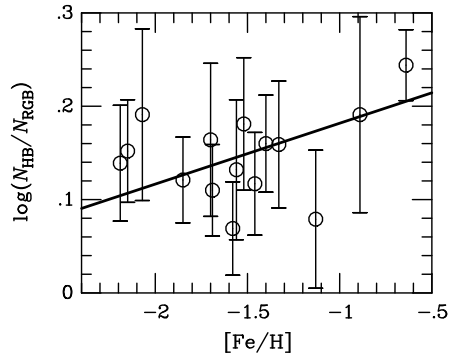}\hss}
\caption{\label{fighbnumber} Number ratio of HB/RGB stars for 15
galactic globular clusters as a function of 
metallicity~\protect\cite{Buzzoni}. The thick line is a linear
fit.}
\end{figure}

The number ratio of HB/RGB stars in 15 galactic globular clusters was
measured in 1983~\cite{Buzzoni}, yielding the results shown in
Fig.~\ref{fighbnumber} as a function of the logarithmic metallicity
measure [Fe/H]. These data reveal that the duration of the
helium-burning phase agrees within about 10\% with the predictions of
the standard stellar evolution theory, leading to the
bound~\cite{RaffeltBook}
\begin{equation}\label{eq:globularlimit}
g_{a\gamma}\alt 0.6\times10^{-10}~{\rm GeV}^{-1}.
\end{equation}
This limit is slightly more restrictive than the often-quoted
``red-giant bound'' which was based on the same argument applied to
open rather than globular clusters~\cite{Dearborn87}. Open clusters
are galactic-disk counterparts to globular clusters; they are less
populous, leading to sparse number counts and hence to statistically
less significant axion limits.  Open clusters tend to have higher
metallicities. As a result, the hydrogen-burning stars are not spread
out along a horizontal branch, but occupy a common location, the
``red-giant clump'' at the base of the RGB. Hence the notion of the
``red-giant bound'' even though the argument as presented here is more
appropriately called ``globular-cluster bound.''

In terms of the Peccei-Quinn scale $f_a$ or axion mass $m_a$, the
axion-photon coupling is
\begin{equation}
g_{a\gamma}=-\frac{3\alpha}{8\pi f_a}\,\xi=
-\frac{m_a/{\rm eV}}{0.69\times10^{10}~{\rm GeV}}\,\xi,
\end{equation}
where
\begin{equation}\label{eq:photoncoupling}
\xi\equiv\frac{4}{3}\left(\frac{E}{N}-1.92\pm0.08\right).
\end{equation}
Here, $E/N$ is a model-dependent ratio of small integers. In the DFSZ
model or GUT models one has $E/N=8/3$, corresponding to $\xi\approx1$.
In this case the globular cluster limit implies
\begin{equation}
m_a\alt0.4~{\rm eV}.
\end{equation}
However, one can construct models where $E/N=2$, allowing for a near
or complete cancellation of the axion-photon coupling. Naturally, in
this case there is no globular-cluster limit on $m_a$ or $f_a$.

One can use other observables in the color-magnitude diagram to test
the theory of stellar evolution, such as the brightness difference
between the HB and the tip of the RGB, the absolute brightness of the
HB, and others. 

In this way one can also derive limits on a putative axion-electron
interaction. It is of the form
\begin{equation}
{\cal L}_{ae}=\frac{C_e}{2 f_a}\,\overline\psi_e\gamma^\mu\gamma_5
\psi_e\,\partial_\mu a
\end{equation}
which is usually equivalent to the pseudoscalar interaction $-i(C_e
m_e/f_a)\,\overline\psi_e\gamma_5\psi_e\,a$. One finds a limit
to the Yukawa coupling of~\cite{RaffeltWeiss}
\begin{equation}\label{eq:electronlimit}
g_{ae}=C_e m_e/f_a\alt2.5\times10^{-13},
\end{equation}
or $\alpha_{26}\alt0.5$ where $\alpha_{26}=10^{26}\,g_{ae}^2/4\pi$.
However, the existence of such a coupling is not generic to all axion
models.

%%%%%%%%%%%%%%%%%%%%%%%%%%%%%%%%%%%%%%%%%%%%%%%%%%%%%%%%%%%%%%%%%%%%%%
%% Section III %%%%%%%%%%%%%%%%%%%%%%%%%%%%%%%%%%%%%%%%%%%%%%%%%%%%%%%
%%%%%%%%%%%%%%%%%%%%%%%%%%%%%%%%%%%%%%%%%%%%%%%%%%%%%%%%%%%%%%%%%%%%%%

\section{WHITE DWARFS}

If axions do interact with electrons, they are emitted from white
dwarfs predominantly by the bremsstrahlung process $e^-+Ze\to
Ze+e^-+a$ and would thus accelerate the cooling of these degenerate
stellar remnants.  The observed white-dwarf luminosity function yields
a limit on $g_{ae}$ which is slightly weaker than
Eq.~(\ref{eq:electronlimit}).

Intruigingly, axion emission with $\alpha_{26}\approx 0.45$, just
below the globular-cluster limit, might dominate the cooling of white
dwarfs such as the ZZ~Ceti star G117--B15A for which the cooling speed
has been established by a direct measurement of the decrease of its
pulsation period~\cite{Isern}.

The most popular example where axions couple to electrons is the DFSZ
model where $C_e={1\over3}\cos^2\beta$ with $\beta$ an arbitrary
angle.  In this case one may use the SN 1987A limits on the
axion-nucleon coupling (see below) to derive an indirect
$\beta$-dependent limit on $g_{ae}$.  From Ref.~\cite{Keil} I infer
that the largest axion-electron coupling allowed by SN~1987A
corresponds to $\alpha_{26}\approx 0.08$, suggesting that DFSZ axions
are not responsible for the anomalous cooling speed of G117--B15A.
 
In magnetic white dwarfs, axion emission can be accelerated by the
cyclotron process~\cite{Kachelriess}.  However, it appears that one
needs unreasonably strong $B$ fields to obtain a significant effect.

%%%%%%%%%%%%%%%%%%%%%%%%%%%%%%%%%%%%%%%%%%%%%%%%%%%%%%%%%%%%%%%%%%%%%%
%% Section IV %%%%%%%%%%%%%%%%%%%%%%%%%%%%%%%%%%%%%%%%%%%%%%%%%%%%%%%%
%%%%%%%%%%%%%%%%%%%%%%%%%%%%%%%%%%%%%%%%%%%%%%%%%%%%%%%%%%%%%%%%%%%%%%

\section{SUPERNOVA 1987A}

\smallskip

Being a QCD phenomenon, axions generically couple to nucleons.  The
most significant limit on this coupling arises from the cooling speed
of nascent neutron stars as established by the duration of the
neutrino signal from the supernova (SN) 1987A. It also implies the
most restrictive constraints on $f_a$ and $m_a$.

A type~II supernova explosion is physically the implosion of the
degenerate iron core of a massive star which has gone through all
possible burning phases. As iron is the most tightly bound nucleus,
further energy gain by nuclear fusion is not possible so that the core
becomes unstable when it has reached the limiting mass (Chandrasekhar
mass) that can be supported by electron degeneracy pressure. The
ensuing collapse is intercepted when the equation of state (EOS)
stiffens at nuclear density, i.e.\ the core mass of 1--$2\,M_\odot$
collapses to the size of a few ten kilometers across.  At nuclear
densities ($\rho_0=3\times10^{14}~{\rm g~cm^{-3}}$) and temperatures
of tens of MeV, this compact object is opaque to neutrinos which are
thus trapped. Therefore, the gravitational binding energy of the newly
formed neutron star (``proto neutron star'') of about
$3\times10^{53}~{\rm ergs}$ is radiated over several seconds from the
``neutrino sphere.'' Crudely put, the collapsed SN core cools by
blackbody neutrino emission from its surface.

\begin{figure}[ht]
\hbox to\hsize{\hss\epsfxsize=6cm\epsfbox{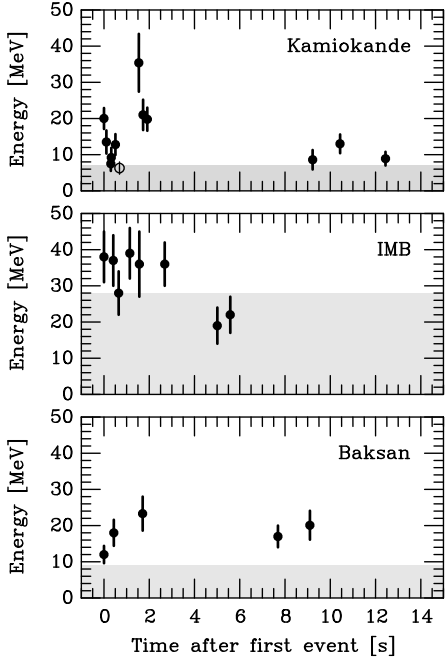}\hss}
\caption{\label{figsndat} SN~1987A neutrino observations at
  Kamiokande~\protect\cite{snkam}, IMB~\protect\cite{snimb} and
  Baksan~\protect\cite{snbst}.  The energies refer to the secondary
  positrons from the reaction $\bar\nu_e p\to n e^+$, not the primary
  neutrinos.  In the shaded area the trigger efficiency is less than
  30\%. The clocks have unknown relative offsets; in each case the
  first event was shifted to $t=0$. In Kamiokande, the event marked as
  an open circle is attributed to background.}
\end{figure}

The neutrinos from a collapsed star were observed once, on 23
Feb.~1987, when the blue supergiant Sanduleak $-69\,202$ in the Large
Magellanic Cloud (a small satellite galaxy of our Milky Way at a
distance of about 165,000 light years) exploded---the legendary
SN~1987A. The neutrinos were registered by the Kamiokande~\cite{snkam}
and IMB~\cite{snimb} water Cherenkov detectors and the Baksan
Scintillator Telescope (BST)~\cite{snbst}---see Fig.~\ref{figsndat}
for a summary of the data. The number of events, their energies, and
the distribution over several seconds corresponds well to what is
theoretically expected and has thus been taken as a confirmation of
the picture that in a type~II supernova a compact remnant forms which
emits its energy by quasi-thermal neutrino emission.

\begin{figure}[t]
\hbox to\hsize{\hss\epsfxsize=2.5cm\epsfbox{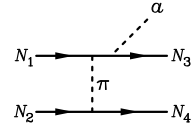}\hss}
\caption{\label{figbremsax} Nucleon-nucleon
bremsstrahlung emission of axions.}
\end{figure}

In analogy to the energy-loss argument for normal stars, the emission
of axions would compete with the standard cooling channel, in the
present case neutrinos, and would thus remove energy from the 
neutrino signal. Axions would be produced by nucleon-nucleon
bremsstrahlung (Fig.~\ref{figbremsax}) in the inner SN core and would
escape directly from there. Therefore, they would primarily remove the
energy which powers the late-time neutrino emission so that their main
effect would be to shorten the expected neutrino signal, in contrast
with the observations~\cite{RS88,Turner88}.

Of course, if the axion coupling were too strong, then they would be
trapped like neutrinos and emerge as thermal radiation from an
appropriate ``axiosphere'' near the surface of the proto neutron
star~\cite{Turner88}.  Therefore, the existence of axions is
compatible with the SN~1987A neutrino signal if axions are either very
weakly or relatively strongly interacting. 

\begin{figure}[ht]
\hbox to\hsize{\hss\epsfxsize=6.1cm\epsfbox{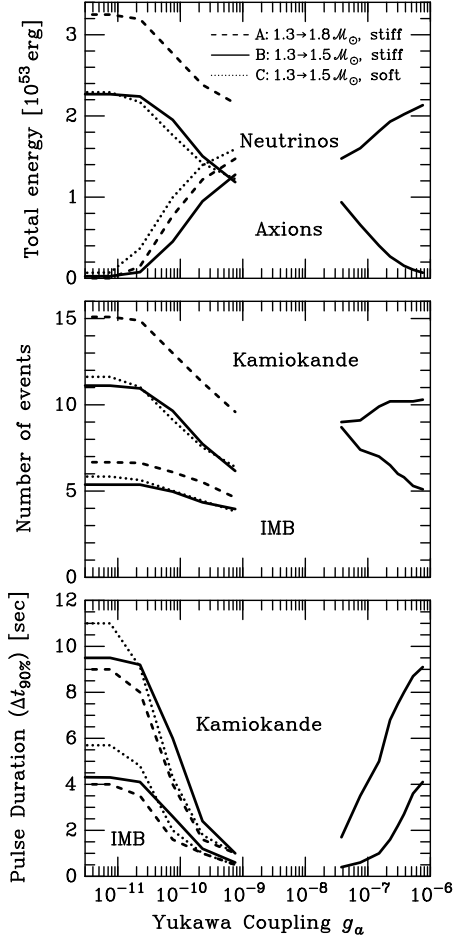}\hss}
\caption{\label{figsnax} Results from numerical protoneutron star
cooling sequences with axions. The free-streaming regime (small $g_a$)
is taken from Ref.~\protect\cite{BTB}, the trapping regime (large
$g_a$) from Ref.~\protect\cite{BRT}. For models A, B, and C
(corresponding to models 57, 55, and 62 of
Ref.~\protect\cite{Burrows88}) the amount of early accretion and the
equation of state (``stiff'' or ``soft'') is indicated. The models
were calculated until 20~s after collapse.}
\end{figure}

This general insight was backed up by detailed numerical simulations,
some of which are summarized in Fig.~\ref{figsnax}. The total energy
emitted in axions and neutrinos, the total number of neutrino events
expected in the Kamiokande and IMB detectors, and the
signal duration are each shown as a function of the
assumed axion-nucleon Yukawa coupling which was taken to be the
same for protons and neutrons. The interaction has the general 
derivative structure
\begin{equation}\label{eq:nucleoncoupling}
{\cal L}_{aN}=\frac{C_N}{2 f_a}\,\overline\psi_N\gamma^\mu\gamma_5
\psi_N\,\partial_\mu a.
\end{equation}
The Yukawa coupling is $g_{aN}\equiv C_N m_N/f_a$ in analogy to the
electron coupling. In general the dimensionless numerical coefficients
$C_N$ will be different for protons and neutrons.  From
Fig.~\ref{figsnax} it is evident that for a certain window of coupling
constants the neutrino signal would be significantly shortened.

It must be stressed that axions on the ``strong interaction'' side of
this window are not necessarily allowed because they themselves could
be registered in the water Cherenkov detectors as they could be
absorbed by nuclei which can then de-excite by $\gamma$
emission~\cite{Engel}.

\begin{figure}[t]
\hbox to\hsize{\hss\epsfxsize=6cm\epsfbox{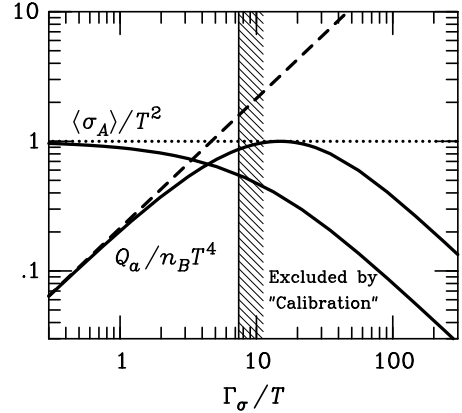}\hss}
\caption{\label{figsuppression} Axion emission rate and 
  thermally averaged neutrino scattering cross section as a function
  of the assumed spin fluctuation rate in a nuclear
  medium~\protect\cite{JKRS}.  The dotted and dashed lines represent
  the ``naive'' perturbative behavior. The duration of the SN~1987A
  neutrino signal excludes $\Gamma_\sigma$ to the
  right of the hatched band.}
\end{figure}

A significant problem with the supernova argument is the difficulty of
calculating the axion emission rate in a hot and dense nuclear medium.
The early bremsstrahlung calculations were based on a ``naive''
evaluation of the perturbative amplitude of Fig.~\ref{figbremsax}.
However, when one looks at this problem from the more general
perspective of linear-response theory one can easily show that such a
calculation is not self-consistent. The derivative coupling of
Eq.~(\ref{eq:nucleoncoupling}) implies that in the nonrelativistic
limit axions couple to the nucleon spins. Therefore, what emits the
axions are the nucleon spins which themselves are being kicked around
by spin-dependent forces among each other. Therefore, the
spin-fluctuation rate $\Gamma_\sigma$, roughly the rate at which a
nucleon spin is flipped in collisions, is an intuitive measure of the
microscopic event rate that leads to the emission of axions. On rather
general grounds one can show that the axion emission rate per nucleon
as a function of $\Gamma_\sigma$ first increases, but later it turns
over and actually decreases as indicated in Fig.~\ref{figsuppression},
a behavior which is naturally understood in terms of multiple
scatterings in the spirit of the Landau-Pomeranchuk-Migdal
effect~\cite{JKRS}.

\begin{figure}[t]
\hbox to\hsize{\hss\epsfxsize=5.6cm
\epsfbox{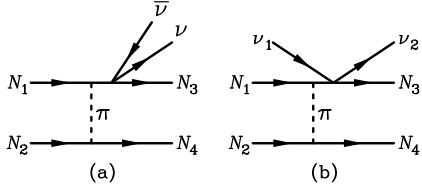}\hss}
\caption{\label{figbremsgraph} Nucleon
bremsstrahlung emission of neutrino pairs and the corresponding
``inelastic scattering'' amplitude.}
\end{figure}

A perturbative calculation of $\Gamma_\sigma$, based on a one-pion
exchange potential, reveals that $\Gamma_\sigma/T$ is 30--50 for the
conditions of a SN core. Such large values likely are significant
overestimates. An empirical reason for this conclusion is based on the
SN~1987A signal duration.  To this end observe that neutrinos couple
to nucleons by an axial-vector interaction. Their vector-current
contribution to scattering cross sections is smaller than the
axial-current contribution by a factor of a few so that we may ignore
it at a first crude level of approximation.  Then it is clear that
neutrino pairs can be emitted in bremsstrahlung processes in full
analogy to axions, but also that spin fluctuations will affect the
neutrino elastic scattering cross section (Fig.~\ref{figbremsgraph}).
One can again show on the basis of general linear-response theory
arguments that this effect leads to a reduction of the axial-current
scattering cross section. Graphically this may be pictured as a
reduction of the nucleon ``average spin'' seen by neutrinos due to
spin-flipping nucleon-nucleon interactions. The expected average
cross-section reduction as a function of $\Gamma_\sigma$ is shown in
Fig.~\ref{figsuppression}.  A large neutrino cross-section reduction
would make the proto neutron star more transparent to neutrinos and
thus would allow them to escape more quickly, unduly shortening the
SN~1987A signal. This argument suggests that values of $\Gamma_\sigma$
to right of the hatched band in Fig.~\ref{figsuppression} are
excluded.

A quantitative understanding of both axion emissivities and neutrino
opacities in a hot and dense nuclear medium requires a reliable
calculation of the dynamical spin and isospin structure functions, a
task which is currently out of reach. It is possible, and even
expected, that nucleon correlations beyond the above simple
multiple-scattering effects provide significant modifications of the
neutrino opacities~\cite{Sawyer} and thus also of the axion
emissivities.

Unless there is some huge unexpected cancellation, however, the axion
emissivities likely are near the maximum indicated in
Fig.~\ref{figsuppression}. Based on this assumption, limits on the
axion-nucleon coupling have been both estimated~\cite{JKRS} and
calculated from detailed numerical cooling sequences~\cite{Keil} which
are not very different from the ones shown in Fig.~\ref{figsnax}
except that the axion emission rate is suppressed according to the
schematic picture of Fig~\ref{figsuppression}.  

In order to translate the limits on the axion-nucleon couplings into
limits on the Peccei-Quinn scale or axion mass one must use specific
models. For KSVZ axions, where the axion-neutron coupling nearly
vanishes, the SN~1987A limit is $m_A\alt0.008\,{\rm eV}$, while it
varies between about 0.004 and $0.012\,{\rm eV}$ for DFSZ axions,
depending on the angle $\beta$ which measures the ratio of two Higgs
vacuum expectation values~\cite{Keil}.  In view of the large overall
uncertainties it is good enough to remember $m_A\alt 0.01\,\rm eV$ as
a generic limit.

%%%%%%%%%%%%%%%%%%%%%%%%%%%%%%%%%%%%%%%%%%%%%%%%%%%%%%%%%%%%%%%%%%%%%%
%% Section V %%%%%%%%%%%%%%%%%%%%%%%%%%%%%%%%%%%%%%%%%%%%%%%%%%%%%%%%%
%%%%%%%%%%%%%%%%%%%%%%%%%%%%%%%%%%%%%%%%%%%%%%%%%%%%%%%%%%%%%%%%%%%%%%

\begin{figure}[b]
\vskip21pt
\hbox to \hsize{\hfil\epsfxsize=\hsize\epsfbox{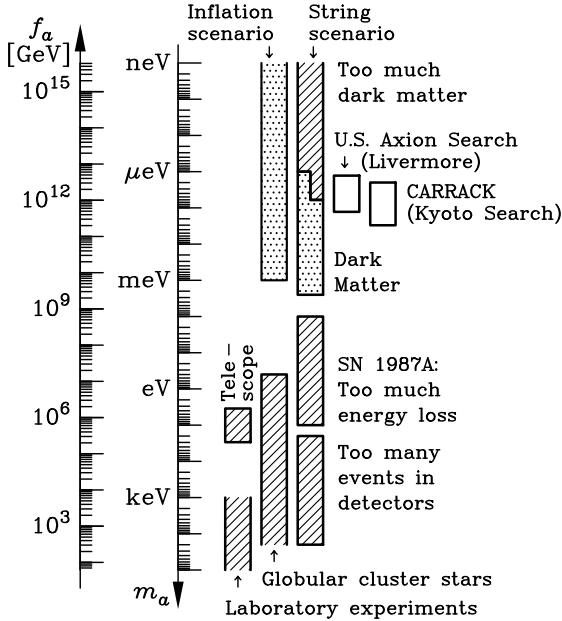}\hfil}
\caption{\label{figaxlim} Astrophysical and cosmological exclusion
regions (hatched) for the axion mass $m_a$ or equivalently, the
Peccei-Quinn scale $f_a$. An ``open end'' of an exclusion bar means
that it represents a rough estimate; its exact location has not been
established or it depends on detailed model assumptions.  The globular
cluster limit depends on the axion-photon coupling; it was assumed
that $E/N=8/3$ as in GUT models or the DFSZ model.  The SN 1987A
limits depend on the axion-nucleon couplings; the shown case
corresponds to the KSVZ model and approximately to the DFSZ model.
The dotted ``inclusion regions'' indicate where axions could plausibly
be the cosmic dark matter.  Most of the allowed range in the inflation
scenario requires fine-tuned initial conditions.  In the string
scenario the plausible dark-matter range is controversial as indicated
by the step in the low-mass end of the ``inclusion bar.''  Also shown
is the projected sensitivity range for the galactic dark-matter search
experiments.}
\end{figure}

\section{COSMOLOGY}

While axion cosmology is not the topic of this overview (various
aspects are covered by other speakers at this conference), a few
remarks are in order to explain my summary plot Fig.~\ref{figaxlim}.

In the early universe, axions come into thermal equilibrium if
$f_a\alt 10^8\,{\rm GeV}$~\cite{Turner87}, a region excluded by the
stellar-evolution limits.  If inflation occurred after the
Peccei-Quinn symmetry breaking or if $T_{\rm reheat}<f_a$, the
``misalignment mechanism''~\cite{Misalignment} leads to a contribution
to the cosmic critical density of $\Omega_a h^2\approx 1.9\times
3^{\pm1} \,(1\,\mu{\rm eV}/m_a)^{1.175}\, \Theta_{\rm i}^2
F(\Theta_{\rm i})$ where $h$ is the Hubble constant in units of
$100\,\rm km\,s^{-1}\,Mpc^{-1}$. The stated range reflects
uncertainties of the cosmic conditions at the QCD phase transition and
of the temperature-dependent axion mass. The function $F(\Theta)$ with
$F(0)=1$ and $F(\pi)=\infty$ accounts for anharmonic corrections to
the axion potential. Because the initial misalignment angle
$\Theta_{\rm i}$ can be very small or very close to $\pi$, there is no
real prediction for the mass of dark-matter axions even though one
would expect $\Theta_{\rm i}^2 F(\Theta_{\rm i})\sim1$ to avoid
fine-tuning the initial~conditions.

A possible fine-tuning of $\Theta_{\rm i}$ is limited by
inflation-induced quantum fluctuations which in turn lead to
temperature fluctuations of the cosmic microwave
background~\cite{Turner91,BS98}. In a broad class of inflationary
models one thus finds an upper limit to $m_a$ where axions could be
the dark matter.  According to the most recent discussion~\cite{BS98}
it is about $10^{-3}~\rm eV$ (Fig.~\ref{figaxlim}).

If inflation did not occur at all or if it occurred before the
Peccei-Quinn symmetry breaking with $T_{\rm reheat}>f_a$, cosmic axion
strings form by the Kibble mechanism~\cite{Davis}.  Their motion is
damped primarily by axion emission rather than gravitational waves.
After axions acquire a mass at the QCD phase transition they quickly
become nonrelativistic and thus form a cold dark matter component. The
axion density such produced is similar to that from the misalignment
mechanism for $\Theta_{\rm i}={\cal O}(1)$, but in detail the
calculations are difficult and somewhat controversial between two
groups of authors~\cite{BattyeShellard,SikivieHagmann}.  Taking into
account the uncertainty in various cosmological parameters one arrives
at a plausible range for dark-matter axions as indicated in
Fig.~\ref{figaxlim}.

If axions are indeed the dark matter of our galaxy one can search for
them with the ``haloscope method.''  At the present time two
full-scale ``second generation'' axion haloscopes are in operation,
one in Livermore (California)~\cite{Livermore} and one in Kyoto
(Japan)~\cite{Kyoto}, the latter one using a beam of Rydberg atoms as
a low-noise microwave detector. The projected sensitivity range shown
in Fig.~\ref{figaxlim} covers the lower end of the plausible mass
range for dark-matter axions.

%%%%%%%%%%%%%%%%%%%%%%%%%%%%%%%%%%%%%%%%%%%%%%%%%%%%%%%%%%%%%%%%%%%%%%
%% Section VI %%%%%%%%%%%%%%%%%%%%%%%%%%%%%%%%%%%%%%%%%%%%%%%%%%%%%%%%
%%%%%%%%%%%%%%%%%%%%%%%%%%%%%%%%%%%%%%%%%%%%%%%%%%%%%%%%%%%%%%%%%%%%%%

\section{SUMMARY}

By virtue of their close relationship to neutral pions, axions couple
generically to photons and nucleons, allowing for the production of
axions in the hot and dense interior of various types of stars. Number
counts of globular-cluster stars reveal that the duration of the
advanced evolutionary phases of these low-mass stars agrees well with
standard predictions, excluding the operation of a strong anomalous
energy-loss channel.  The SN~1987A neutrino signal further indicates
that axions with a mass exceeding about $10^{-2}~{\rm eV}$ would
unduly shorten the cooling time of a proto neutron star. Taken
together, these limits suggest that axions, if they exist, play a
cosmological dark-matter role, at least in the framework of a scenario
where primordial axions are produced by string radiation rather than
the misalignment mechanism.

One interesting loop-hole arises in axion models where the coupling to
photons is very small because $E/N=2$ in
Eq.~(\ref{eq:photoncoupling}), leading to a cancellation effect. In
this case there is a narrow window of allowed axion masses around a
few eV on the strong-interaction side of the SN~1987A limit.
Intruigingly, axions in this window would be important as a hot dark
matter component~\cite{Moroi}.

From the perspective of stellar-evolution tests of the axion
hypothesis, it remains to be hoped that we will be lucky and see a
galactic supernova with a large detector such as Superkamiokande to
obtain a high-statistics neutrino signal. Meanwhile, more work is
needed on the microscopic input physics for numerical supernova
studies, both for the sake of a better understanding of the supernova
phenomenon and its application as a laboratory for fundamental
physics.

%%%%%%%%%%%%%%%%%%%%%%%%%%%%%%%%%%%%%%%%%%%%%%%%%%%%%%%%%%%%%%%%%%%%%%
%% Acknowledgments %%%%%%%%%%%%%%%%%%%%%%%%%%%%%%%%%%%%%%%%%%%%%%%%%%%
%%%%%%%%%%%%%%%%%%%%%%%%%%%%%%%%%%%%%%%%%%%%%%%%%%%%%%%%%%%%%%%%%%%%%%

\section*{ACKNOWLEDGMENTS}

Partial support by the Deutsche Forschungsgemeinschaft under
grant No.\ SFB-375 is acknowledged.

\bigskip

%%%%%%%%%%%%%%%%%%%%%%%%%%%%%%%%%%%%%%%%%%%%%%%%%%%%%%%%%%%%%%%%%%%%%%
%% References %%%%%%%%%%%%%%%%%%%%%%%%%%%%%%%%%%%%%%%%%%%%%%%%%%%%%%%%
%%%%%%%%%%%%%%%%%%%%%%%%%%%%%%%%%%%%%%%%%%%%%%%%%%%%%%%%%%%%%%%%%%%%%%

\end{document}